\documentclass{emulateapj}
\usepackage{graphicx}
\usepackage{amsmath}
\usepackage{amssymb}
\usepackage{verbatim}
\usepackage{longtable}
\usepackage{url}
\shorttitle{Optical Structure of Low Mass Galaxies at $2 \lesssim z \lesssim 3$}
\shortauthors{Hagen et al.}

\def\etal{{et~al.\null}}

\begin{document}

\title{The Rest-Frame Optical Morphology of Emission Line Galaxies at 
$2 \lesssim \MakeLowercase{z} \lesssim 3$:\\  Evidence for Inside-Out Galaxy Formation in Low-Mass Galaxies}
\author{Alex Hagen\altaffilmark{1, 2}}  \author{Nicholas A. Bond\altaffilmark{3}}
\author{Robin Ciardullo\altaffilmark{1,2}} \author{Caryl Gronwall\altaffilmark{1,2}}
\author{Eric Gawiser\altaffilmark{4}} \author{William Bowman\altaffilmark{1, 2}} \author{Joanna S. Bridge\altaffilmark{1, 2}}
\author{Henry S. Grasshorn Gebhardt\altaffilmark{1, 2}} \author{Donald P. Schneider\altaffilmark{1, 2}}
\submitted{ApJ, Submitted}
\altaffiltext{1}{Department of Astronomy and Astrophysics, Pennsylvania State University, University 
Park, PA 16802, USA}
\altaffiltext{2}{Institute for Gravitation and the Cosmos, Pennsylvania State University, 
University Park, PA 16802, USA}
\altaffiltext{3}{Cosmology Laboratory (Code 665), NASA Goddard Space Flight Center, Greenbelt, 
MD 20771, USA}
\altaffiltext{4}{Physics and Astronomy Department, Rutgers University, Piscataway, NJ 
08854, U.S.A.}

\begin{abstract}
We compare the rest-frame ultraviolet and rest-frame optical morphologies of 
$2 \lesssim z \lesssim 3$ star-forming galaxies in the GOODS-S field using {\sl
Hubble Space Telescope\/} WFC3 and ACS images from the CANDELS, GOODS, and ERS
programs. We show that the distribution of sizes and concentrations for $1.90 < z < 2.35$ 
galaxies selected via their rest-frame optical emission-lines are statistically indistinguishable 
from those of Ly$\alpha$ emitting systems found at $z \sim 2.1$ and $z \sim 3.1$.   We also show 
that the $z \gtrsim 2$ star-forming systems of all sizes and masses become smaller and more 
compact as one shifts the observing window from the UV to the optical. We argue that
this offset is due to inside-out galaxy formation over the first $\sim 2$~Gyr of cosmic time.

\end{abstract}


\vspace{0.4in}

\section{Introduction}

The majority of galaxies in the local universe lie on the Hubble sequence \citep{HubbleSeq}, a 
continuum that runs from red, passively-evolving compact ellipticals to gas-rich, 
star-forming disks with exponential surface brightness profiles.  This morphological sequence 
is clearly visible out to intermediate redshifts \citep[e.g.,][]{Glazebrook+95, vdB+96, 
Griffiths+96, Brinchmann+98, Lilly+98, Simard+99, vanDokkum+00, Stanford+04, 
Ravindranath+04}, but by $z \sim 2$ the relationship breaks down 
\citep[e.g.,][]{Giavalisco+96, Lowenthal+97, Dickinson00, vdB01, Papovich+05, Conselice+05,
Pirzkal+07}.   In fact, a number of surveys have demonstrated that most star-forming galaxies 
between $2 \lesssim z \lesssim 3$ are clumpy,  disturbed, or disk-like, and only $\sim 30\%$ 
have surface brightness profiles consistent with galactic spheroids \citep[e.g.,][]{Ferguson+04,
Elmegreen+05, Lotz+06, Ravindranath+06, Petty+09, Tacchella+15}.  These studies also show
that the typical half-light radii of star-forming galaxies at $z \gtrsim 2$ is $\sim 2$~kpc,  
and that their sizes evolve approximately as $H^{-1}(z)$. 

There are two limitations to the analyses cited above.  The first is associated with the
wavelength of observation.  To date, most morphological surveys of 
$z \gtrsim 2$ galaxies have been conducted in the rest-frame ultraviolet, where light from 
newly-born stars dominates.   This limitation can have a profound effect on the results, as in the 
nearby universe, galaxies become less concentrated, more clumpy, and more asymmetric 
as one shifts the observing window from the red and infrared to the ultraviolet 
\citep[e.g.,][]{Taylor-Mager+07}.   Consequently, to properly compare the morphology of 
$z \gtrsim 2$ galaxies to systems in the local universe, one must move from the
rest-frame UV to (at least) the rest-frame optical. Only recently have there been
investigations of this type \citep[e.g.,][]{Bond+11, Patel+13, Lang+14, vanderwel+14}.

A second limitation of most morphological analyses arises from the galaxy sample
selection.  Most studies of $z \gtrsim 2$ systems have focused on objects that
were identified, at least in part, on the basis of their continuum brightness.  This criterion 
creates a bias in favor of high stellar-mass objects located on the bright end of galaxy 
luminosity function.  Lower-mass objects, which represent the bulk of the epoch's galaxies, 
have not been well-represented.

In this investigation, we address these two problems using images from the
 \textit{Hubble Space Telescope's (HST)} Wide Field Camera 3 (WFC3) and the Advanced Camera 
for Surveys (ACS) that were taken as part of the Cosmic Assembly Near-IR Deep Extragalactic Legacy 
Survey (CANDELS), the Great Observatories Origin Deep Survey (GOODS), and the WFC3 Early 
Release Science (ERS) Program \citep{GOODS, wfc3ers, Grogin+11, Koekemoer+11}.   Our targets for study 
are $2 \lesssim z \lesssim 3$ star-forming galaxies in the GOODS-S Field, 
which have been identified via their emission lines, either in the rest-frame optical (oELGs) 
or through Ly$\alpha$ (LAEs).  
Because the multi-wavelength \textit{HST} surveys of this region 
extend from the UV through the near-IR and reach as deep as $m_{AB} \sim 26.4$ in the 
F160W ($H$) band, we can probe the morphology of $z \gtrsim 2$ systems in the 
rest-frame optical, and compare their structure to that measured in the rest-frame UV\null. 
Moreover, by analyzing sets of emission-line galaxies drawn from the entire range of stellar 
masses, we can examine how the systematics of morphology depend on mass.

In \S\ref{sec:data} we begin by describing the galaxy samples used in this study.  
In \S\ref{sec:images} and \ref{sec:method}, we summarize the properties of the
images and describe the analysis techniques used in our comparative 
study.  In \S\ref{sec:results}, we present the rest-frame optical sizes and concentration
indices of our galaxies, and compare these data to similar measurements made at shorter 
wavelengths. In \S\ref{sec:discussion}, we discuss the implications of our findings. 

Throughout this paper, we assume a concordance cosmology with 
$H_0=70$~km~s$^{-1}$~Mpc$^{-1}$,  $\Omega_{\rm M}=0.3$, and $\Omega_{\Lambda}=0.7$
 \citep{WMAP}.  With these values, $1\arcsec = 8.320$~kpc at $z=2.1$ and 7.625~kpc at 
 $z=3.1$.
 
\section{The Galaxy Samples}
\label{sec:data}
The first step towards understanding the morphological systematics of $z \gtrsim 2$ 
star-forming galaxies is to select a sample of objects representative of the epoch.   This 
process is not straightforward.  High-redshift galaxies identified on the basis of their rest-frame
UV or optical continuum brightness \citep[i.e., Lyman-break and BzK 
galaxies;][]{Giavalisco02, Daddi+04} or thermal dust 
properties \citep[{\sl Herschel\/} PACS objects, e.g.,][]{Rodighiero+11} are generally 
high-mass systems destined to become 
today's giant ellipticals \citep[e.g.,][]{Adelberger+05, Quadri+07}.    The best
way to reach galaxies further down the mass function is to identify systems from their
emission lines, which are excited either by recombination or collisions.  In particular,
studies of $z \gtrsim 2$ star-forming galaxies have demonstrated that objects
selected on the basis of their Ly$\alpha$ or [O~III] $\lambda 5007$ line-strength can span
an extremely wide range of stellar mass, from $7.5 \lesssim \log M/M_{\odot} \lesssim 10.5$ 
\citep{Hagen+14, Hagen+16}.  

For our examination of rest-frame optical morphologies, we therefore used three samples of
$z \gtrsim 2$ emission-line galaxies in the GOODS-S field.  The first set of objects consists
of star-forming galaxies identified via their rest-frame optical emission lines, which we call oELGs.  Data from the
WFC3 G141 grism are ideal for this purpose, and are available in GOODS-S as a consequence of
the 3D-HST survey \citep{Brammer+12}.   By visually examining these frames, \citet{Zeimann+14}
identified a sample of 64 objects in GOODS-S with multiple emission lines, unambiguous redshifts
between $1.90 < z < 2.35$, and F125W + F160W continuum magnitudes brighter than $m_{AB} = 26.$
An analysis of the galaxies' spectral energy distributions (SEDs) demonstrates that these objects have 
a very wide range of stellar  masses ($7.5 \lesssim \log M/M_{\odot} \lesssim 10.5$), star 
formation rates ($1 \lesssim M_{\odot}~{\rm yr}^{-1} \lesssim 500$) and stellar reddenings
($0 \lesssim E(B-V)  \lesssim 0.4$) \citep{Hagen+16, Gebhardt+16}.

Our second sample 
is chosen from the set of $z \sim 2.1$ Ly$\alpha$ emitters identified by \citet{Guaita+10} via deep, 
narrow-band [O~II] $\lambda 3727$ images taken with the Mosaic CCD camera on the 
Blanco 4-m telescope.  These data, which are part of the Multiwavelength Survey by Yale-Chile 
\citep[MUSYC;][]{MUSYC}, are defined to have Ly$\alpha$ rest-frame equivalent 
widths EW$_0 > 20$~\AA\ and Ly$\alpha$ line fluxes greater  than $F_{\mathrm{Ly}\alpha} > 
2 \times 10^{-17}$~ergs~cm$^{-2}$~s$^{-1}$, or, equivalently, Ly$\alpha$ emission 
line-luminosities greater than $\log L(\mathrm{Ly}\alpha) > 41.8$~ergs~s$^{-1}$
\citep{Ciardullo+12}.  Originally, \citet{Guaita+10} discovered 250 LAE candidates over a
$\sim 0.3$~deg$^2$ region of the Extended Chandra Deep Field South \citep[ECDF-S;][]{cdfs}; however,
after excluding those objects located outside the range of deep \textit{HST} imaging, positioned 
within 40 pixels of the edge of a CANDELS image, superposed within 2\arcsec\ of a 
cataloged X-ray source \citep{Lehmer+05, Virani+06, Luo+08}, or projected onto a region with an
obvious image defect, this $z \sim 2.1$ sample reduces to a set of 24~objects.   Like the oELGs
described above, this set of LAEs span a very wide range of stellar mass, from 
$\log M/M_{\odot} \sim 7.3$ to $\log M/M_{\odot} \sim 9.5$ 
\citep{vargas+14}.

Finally, to supplement the $z \sim 2$ LAEs, we included two sets of $z \sim 3$ Ly$\alpha$
emitters identified via narrow-band [O~III] $\lambda 5007$ imaging with the Mosaic camera of
the Blanco telescope \citep{Gronwall+07, Ciardullo+12}.  These galaxies, which were also selected 
to have rest-frame Ly$\alpha$ equivalent widths greater than 20~\AA, have slightly larger Ly$\alpha$ 
line luminosities than the $z \sim 2.1$ LAEs ($L({\rm Ly}\alpha) > 42.1$ and $42.3$~ergs~s$^{-1}$ 
for the \citet{Gronwall+07} and \citet{Ciardullo+12} samples, respectively), but roughly the 
same median stellar mass \citep{Gawiser+07, Acquaviva+11}.   A total of 20 of these
 $z \sim 3.1$ galaxies have deep \textit{HST} imaging.

Before analyzing these samples, one additional criterion must be satisfied.  \citet{Bond+09}
has shown that morphological analyses of high-redshift galaxies cannot be performed at the image
sensitivity limit. While \citet{Bond+09} used a signal-to-noise ratio of 30 in their analyses, we adopt a 
minimum signal-to-noise ratio of 10, while keeping track of the increased uncertainties associated with
these lower signal-to-noise measurements.   With this restriction, our samples reduce to 23 LAEs at 
$z \sim 2.1$, 12 LAEs at $z \sim 3.1$, and 61 oELGs with $1.90 < z < 2.35$.   

\section{The Images}
 \label{sec:images}
The source material for our study are the optical through near-IR images taken as part
of the CANDELS \citep{Grogin+11, Koekemoer+11}, GOODS  \citep{GOODS},
and WFC3 ERS programs \citep{wfc3ers}.  Of primary use are the F160W 
frames from the IR arm of WFC3\null.  These $H$-band data, which have a 
(drizzled) image scale of $0\farcs 06$~per pixel and a limiting magnitude of 
$m_{AB} \sim 27.2$, sample the rest-frame optical of our $z \gtrsim 2$ emission-line 
galaxies extremely well.  For $1.90 < z < 2.35$ oELGs, the rest-frame 
wavelengths covered by the filter go from roughly 4800~\AA\ to 5800~\AA\  for galaxies at the blue 
end of the redshift window, and from $\sim 4200$~\AA\ to $\sim 5000$~\AA\ for our 
highest-redshift objects.  For the $z \sim 2.1$ LAEs, the bandpass of the filter extends from 
4500~\AA\ to 5500~\AA, while for the $z \sim 3.1$ LAEs, the range is 
3400~\AA\ $< \lambda < 4100$~\AA\null.  

Supplementing these data are ACS frames covering the galaxies' rest-frame UV continuum.
Specifically, we examined data taken through the F435W, F606W, F775W, F814W, and
F850LP filters, which, for our oELGs and $z \sim 2.1$ LAEs, span the rest-frame
wavelengths from $\sim 1300$~\AA\ to $\sim 3000$~\AA, depending on the exact 
redshift.  For the $z \sim 3.1$~LAEs, the F435W filter is excluded from our analysis,
as it lies blueward of Ly$\alpha$ and may be affected by intervening hydrogen absorption 
\citep{Madau95}.   The remaining filters extend from just redward of Ly$\alpha$ to about
2300~\AA\null.  These ACS data reach depths comparable to that of the
F160W data, but have a better drizzled image scale of $0 \farcs 03$~per pixel.

\section{Measuring Size and Concentration}
\label{sec:method}
The structural properties of our $z \gtrsim 2$ galaxies were measured using a reduction 
pipeline very similar to that developed by \citet{Bond+09} for the analysis of rest-frame 
UV images.   We began with the WFC3 F160W frames and extracted an $18\arcsec \times 
18\arcsec$ postage-stamp region around each galaxy.  We then identified all possible 
sources within each galaxy cutout by using the SExtractor catalog builder  \citep{SExtractor} 
with extraction parameters {\tt DETECT\_MINAREA}$=30$,  {\tt DEBLEND\_MINCONT}
$=0.06$, and a uniform background.   A second SExtractor pass, this time with 
{\tt DETECT\_MINAREA}$=5$, was employed to compute the flux-weighted centroid 
of each object.

We note that in this particular analysis of $z \gtrsim 2$ galaxy morphology, the identification of
F160W counterparts to our emission-line selected galaxies was not an issue.
For the oELGs, which were originally selected using the G141 infrared grism on the
WFC3, the identification of the F160W source was part of the discovery 
and spectral extraction process \citep{Zeimann+14}.  For the LAEs, the formal 
$\sim 0\farcs 25$ astrometric uncertainty of the MUSYC frames was confirmed by 
\citet{Bond+09} during their analysis of LAEs on {\sl HST\/} rest-frame UV images.  With this
precision ($\sim4$ pixels on the archival F160W frames), there is little ambiguity as to the 
most likely counterpart.  Finally, most of the galaxies studied in this program have data in at least
six filters:  F435W, F606W, F775W, F814W, F850LP, an F160W\null.   This wide wavelength
coverage, coupled with the requirement that the galaxy be well-detected (S/N $> 10$) on the
F160W image, ensured that any source present in the discovery image was also detectable 
on multiple CANDELS frames.

Once the likely F160W counterparts were identified, we measured their sizes through a series of 
circular apertures centered on the galaxies' centroids in the F160W image.  (Because of the
relatively high signal-to-noise ratio of the detections, the choice of filter used for this centroiding
has little effect on the analysis.)   We then estimated each galaxy's Petrosian-like radius 
by calculating the radial distance at which the galaxy's local surface brightness, $I(r)$, falls to 
half the mean surface brightness contained with its aperture \citep{Petrosian}.  In other words, 
\begin{equation}
\eta(r) = {I(r) \over \langle I(<r) \rangle} = 0.5
\end{equation}
As this quantity is defined in terms of a surface brightness ratio,
it is relatively insensitive to the depth of the image and thus a robust measure of size.
Moreover, in almost all cases, the  $\eta(r) = 0.5$ radius is very close to 
what would be derived for the galaxy's half-light radius, if the surface brightness profile were 
to be extrapolated to infinity \citep{Bershady+00}.

Along with size, we also computed each galaxy's compactness as viewed in the
rest-frame optical.  Following \citet{Kent85} we defined a system's concentration index 
using the ratio of radii enclosing 80\% and 20\% of the galaxy's light, i.e.,
\begin{equation}
C=5 \log \left[\frac{r_{80\%}}{r_{20\%}}\right]
\end{equation}
Through this definition, elliptical galaxies in the local universe would have values
of $C \sim 5$, bulgeless spiral galaxies would have $C \sim 3$, and a Gaussian
profile would have $C = 2.1$ \citep{Bershady+00}.

Since cosmological surface-brightness dimming makes it difficult to measure the total luminosity
of faint, distant galaxies, we again adapted a dimensionless ratio of surface brightnesses in
our measurements of concentration. 
\citet{Conselice03} demonstrated that for intermediate redshift objects, the total magnitude of a 
galaxy is well approximated by the light contained within a radius that is 1.5 times that defined by
$\eta = 0.2$, and it is this value we use to compute $C$.

Finally, we repeated the above analyses in the rest-frame ultraviolet, using the deep
F435W, F606W, F775W, F814W and F850LP images of the fields.  These data, which
are drizzled to $0\farcs 03$~per pixel, have roughly twice the resolution
as the F160W frames, but reach a similar depth.  These additional images allowed us to trace 
the behavior of concentration and size versus the wavelength of observation.

To estimate the uncertainties on our galaxy sizes, we used the results of
\citet{Bond+12}, who conducted a series of Monte Carlo simulations on the F814W images
of the GOODS \citep{GOODS}, GEMS \citep{GEMS}, and HUDF surveys.
Their analysis showed that the fractional uncertainty in the measured half-light radius
is related to the total flux, $f$, contained within an object by 
\begin{equation}
{\sigma_{r_e} \over r_e} = 0.54 {\sigma_f \over f}
\end{equation}
where $\sigma_f$ is the pixel-to-pixel flux uncertainty derived from the weight map, i.e.,
$\sigma_f = \sqrt{1/W}$.  We adopt this relation for measurements in all our filters.

\section{Results}
\label{sec:results}
\begin{figure}[t]
\plotone{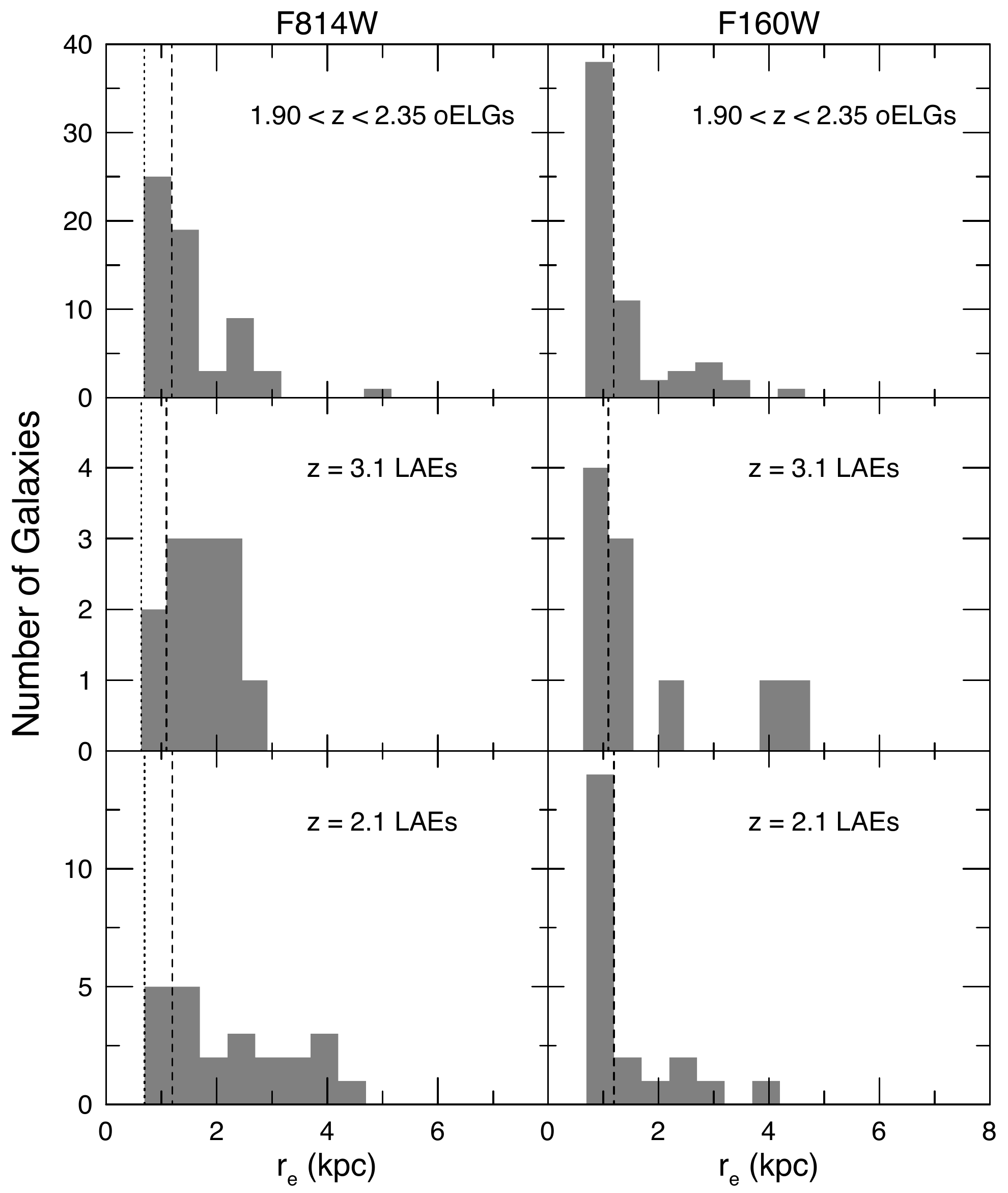}
\caption{Distribution of half-light radii for oELGs and LAEs in the rest frame UV (F814W; left) and 
the rest frame optical (F160W; right).   The dashed (F160W) and dotted (F814W) lines show the limits 
of our ability to resolve objects, given \textit{HST's} PSF and the finite pixel size of the instruments. 
The LAE size distributions are consistent with that found for the oELGs.  Note that 62 out of the 
96 galaxies are unresolved on the F160W frames;  if the F160W sizes were the same as that seen 
through the F814W filter, only 34 of the galaxies would be unresolved.
\label{fig:sizes}}
\end{figure}

Our size and concentration results for oELGs and LAEs in the F814W and F160W filters 
are presented in Tables \ref{tbl:oelg} and \ref{tbl:lae}. 
Figure~\ref{fig:sizes} displays histograms of the half-light radii listed in these tables.  
Before interpreting these data, however, it is important
to understand their limitations.   The majority of the galaxies detected in our survey have
small angular sizes and may be unresolved, even at \textit{HST} resolution.  To investigate this
possibility,  we determined the instrumental resolution of our images using \texttt{Tiny Tim}, 
the \textit{HST's} point-spread-function (PSF) modeling program \citep{tinytim}. 
Using this software, we created a series of point source images through the ACS/F814W and 
WFC3/F160W filters and measured their half-light radii in exactly the same manner as for our 
galaxies.  These simulations reveal that objects with $r_e$ less than 2.8 pixels 
($0\farcs 083$) in the ACS's F814W filter and 2.4 pixels ($0\farcs 14$) in WFC3's F160W 
filter have sizes consistent with that of the median value of our modeled point sources.  
Any galaxy with a value of $r_e$ smaller than these limits should therefore be treated as unresolved.
In Figure~\ref{fig:sizes}, the resolution limits are shown as a dotted line (for the F814W frames)
and a dashed line (for the F160W data).

As is apparent from Figure~\ref{fig:sizes}, virtually all of the $z \sim 2$ star-forming galaxies
present in our sample are (at least marginally) resolved at rest-frame ultraviolet
wavelengths.  However, when viewed at longer wavelengths with the $0 \farcs 06$~per pixel
plate scale of the WFC3's infrared camera, more than half ($\sim 60\%$) of our targets have 
profiles consistent with that of a point source.   For the LAEs, only 12 of the 35 ($\sim 34\%$)
are resolved at WFC3 resolution, while for the oELGs the fraction is 21 out of 61 (again,
$\sim 34\%$).   Although this limitation prevents us from probing the full range of rest-frame optical 
morphologies exhibited by our $z \gtrsim 2$ emission-line sources, we can still use the data
to draw conclusions about the epoch's star-forming population.

The first conclusion arises from the similarity between the size distribution of 
LAEs and that found for galaxies identified via their rest-frame optical emission lines. 
A Kolmogorov-Smirnov (K-S) test can find no statistically significant difference in the distributions, 
either in the UV or in the optical.  The same is true for the galaxies' concentration index, which 
typically falls between 2 and 3: 
the distribution of $C$ values for LAEs is statistically indistinguishable from that of oELGs. 
Although the number of Ly$\alpha$ emitters in our study is relatively 
small, the result does confirm the analysis of \citet{Hagen+16}, who showed that galaxies 
selected on the basis of their Ly$\alpha$ emission have very similar properties to those of
systems found via their rest-frame optical emission lines.  Our concentration indices of LAEs
 in the rest-frame UV also match those of \cite{Gronwall+11}, who measured the same quantity
 in $z \sim 3$ LAEs.

\begin{figure}[t]
\plotone{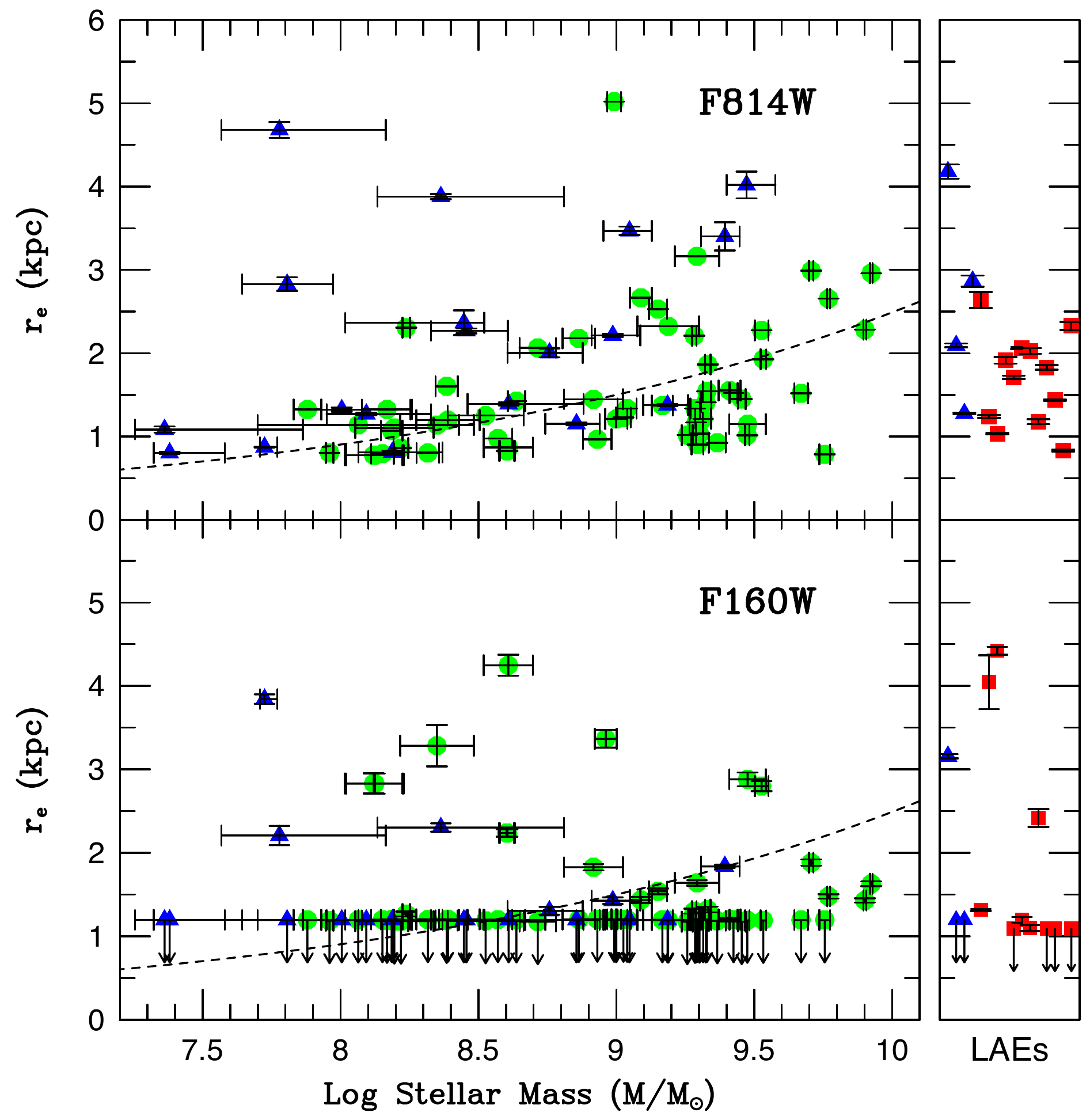}
\caption{The stellar mass-half-light radius relationship for $2 \lesssim z \lesssim 3$ star-forming 
galaxies in the F160W and F814W filters.  The oELGs are green circles, the $z \sim 2$ LAEs are 
blue triangles, and the $z \sim 3.1$ LAEs are shown as red squares. The stellar mass 
measurements come from \cite{vargas+14} and \cite{Hagen+16}; the dashed line shows the
mean size-mass relation derived by \citet{vanderwel+14} for $z \sim 2.25$ star-forming galaxies.
In the rest-frame UV, galaxy size does correlate with stellar mass, but with a large amount of scatter.   
While a size-mass relation may exist in the rest-frame optical, the poorer resolution of the IR
data currently precludes its measurement.  The panels on the right show LAEs without stellar mass 
measurements.
\label{fig:radius_mass}}
\end{figure}

A second result comes from the wide-range of sizes exhibited by our emission-line selected
galaxies.   Although the bulk of the galaxy population is extremely compact, with half-light 
radii less than $\sim 2$~kpc, there is a tail to the distribution that extends to almost 6~kpc.  
Interestingly, this spread is not entirely driven by stellar mass.  As Figure~\ref{fig:radius_mass} 
illustrates, there is a correlation between radius and stellar mass, but the scatter about
the relation is substantial.   At the high-mass end ($M \gtrsim 10^9 M_{\odot}$), our rest-frame UV 
results are consistent with the analysis of \citet{vanderwel+14}, who have traced the $0 < z < 3$ evolution 
of the size-mass distribution using continuum-selected CANDELS galaxies with photometric results.
Our results also agree with the measured the sizes and masses of $z = 2.53$ Ly$\alpha$ and 
H$\alpha$ selected galaxies with $M \gtrsim 10^8 M_{\odot}$ \citep{shimakawa+16}.   Unfortunately, 
while we would like to derive a similar relation for the galaxies in the rest-frame optical,
the limited resolution of the {\sl HST's\/} WFC3 IR camera currently precludes this possibility.

\begin{figure}[t]
\plotone{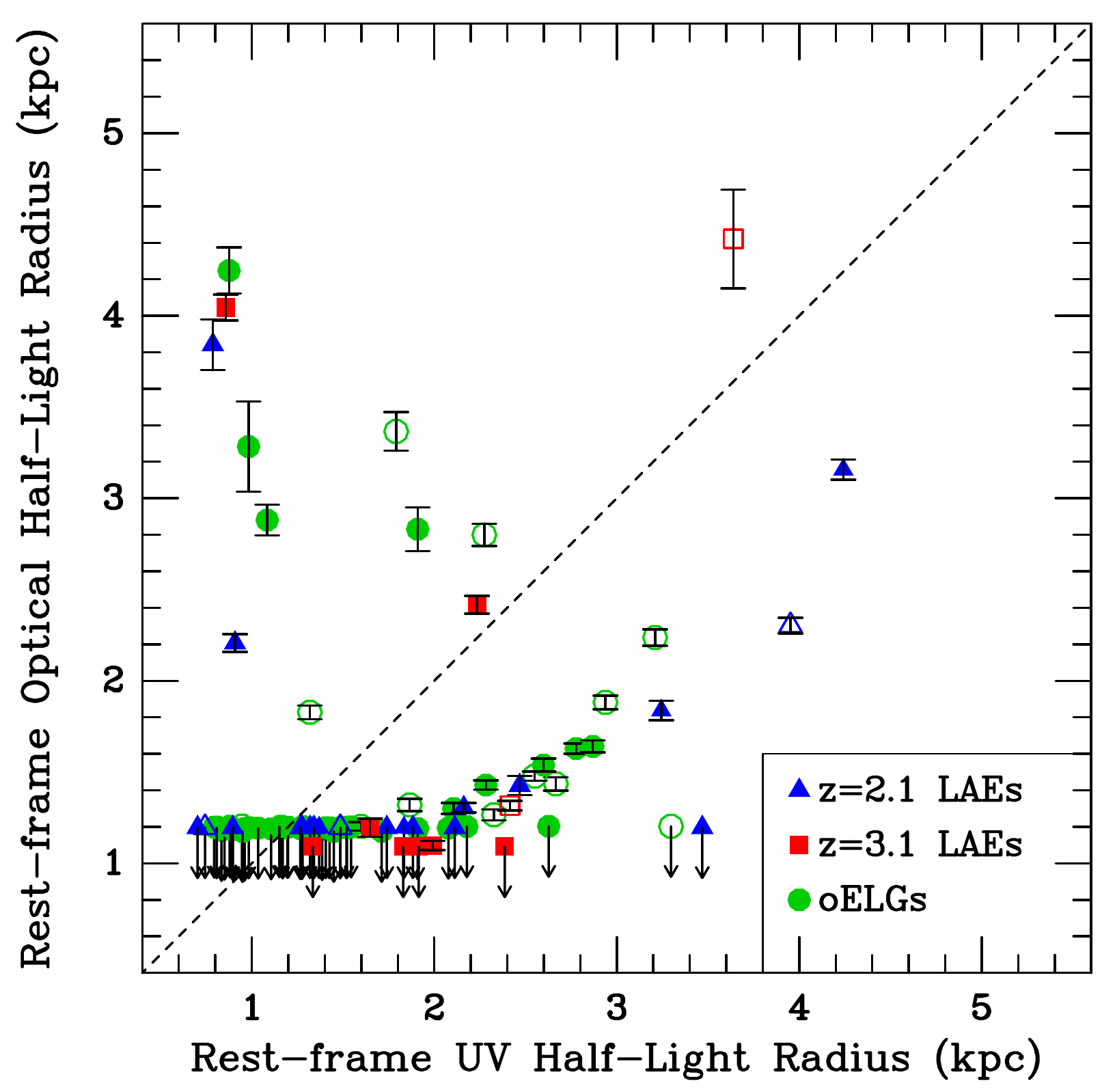}
\caption{The half-light radii of our $z \gtrsim 2$ star-forming galaxies in rest-frame optical 
(WFC3/F160W) vs.\ similar measurements made in the rest-frame UV using ACS images
in F435W, F606W, F775W, F814W, and F850LP\null.  As in
Figure~\ref{fig:radius_mass}, green circles represents the galaxies identified via their rest-frame 
optical emission lines, blue triangles show the $z \sim 2.1$~LAEs, and red squares denote the 
$z \sim 3.1$~LAEs.   The open symbols denote objects with companions projected within 
$0 \farcs 6$.  All unresolved galaxies in F160W are indicated as upper limits at the half-light 
radius of the instrument's PSF\null.  The vast majority of galaxies are smaller in the rest-frame optical 
than the rest-frame UV\null.
\label{fig:RvR}}
\end{figure}

The fact that many of our $z \gtrsim 2$ emission-line galaxies cannot be resolved in the
F160W frames leads to the most important result of this study:  most of the galaxies analyzed
in this program have smaller half-light radii in the rest-frame optical than they do 
in the rest-frame ultraviolet.   This result is displayed vividly in Figure~\ref{fig:RvR}, where
we plotted the rest-frame UV half-light radius of each galaxy, as derived from their median
size on the F435W, F606W, F775W, F814W, and F850LP frames, against their 
their half-light radius in F160W\null.   The offset in the sizes is obvious.  The relation is not 
perfect, and some galaxies are larger in the rest-frame optical than they are in the UV\null.  
Nevertheless, the result is striking, and demonstrates that the star-forming regions detected by
the rest-frame UV images of \textit{HST} are not simply knots imbedded in larger, somewhat
older systems.

\begin{figure}[t]
\plotone{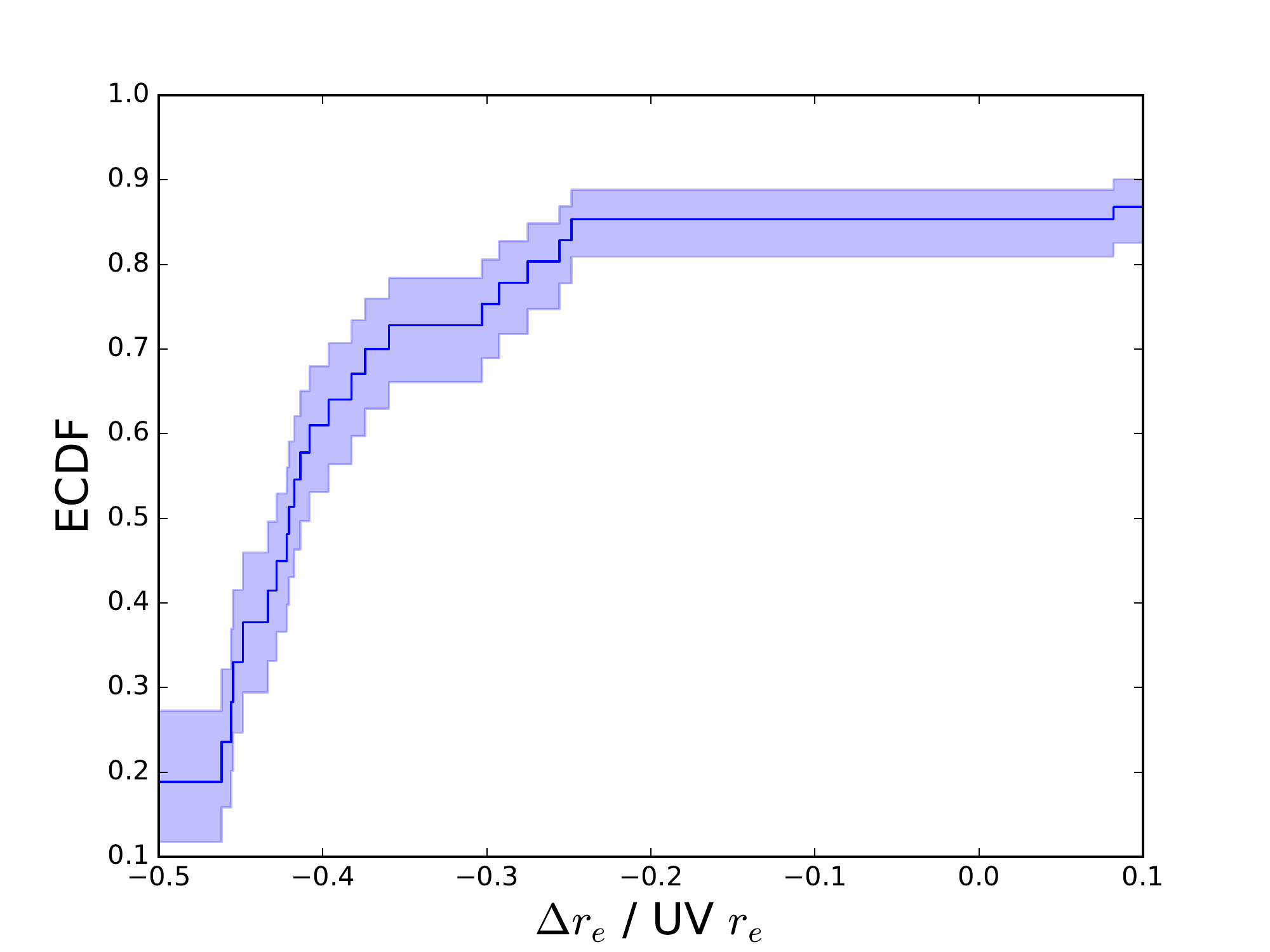}
\caption{The Kaplan-Meier empirical cumulative distribution function (ECDF) for our 
combined sample of $z \gtrsim 2$ oELGs plotted against the fractional change in size
of the galaxy between the rest-frame optical and rest-frame UV, i.e., $\left( r_e(\mathrm{Opt}) - 
r_e(\mathrm{UV}) \right) / r_e(\mathrm{UV})$. The shaded area denotes the $1 \sigma$ 
confidence interval.   The median fractional change between the rest-frame UV and optical is $-0.40$, 
and $\sim 85\%$ of our galaxies are smaller at longer wavelengths.}
\label{fig:kaplan_ecdf}
\end{figure}

The difference between the rest-frame optical and rest-frame UV sizes can be 
quantified using the Kaplan-Meier estimator for the empirical cumulative distribution function 
\citep{kaplan-meier}.   This estimator accounts for the existence of censored data (i.e., points with
only upper limits), and calculates $1 \, \sigma$ confidence intervals on the solution using a method 
developed by \citet{greenwood1926}.  (See Chapter 10 of \citeauthor{msma} for more on information 
on the analysis of censored data.)  The results of the Kaplan-Meier test are shown in 
Figure~\ref{fig:kaplan_ecdf}.  From the figure, it is apparent that $\sim 85\%$ of our oELGs and LAEs 
are smaller in the rest-frame optical than they are in the UV, and that the median fractional change 
in $r_e$ is 40\%.  

\begin{figure}[t]
\plotone{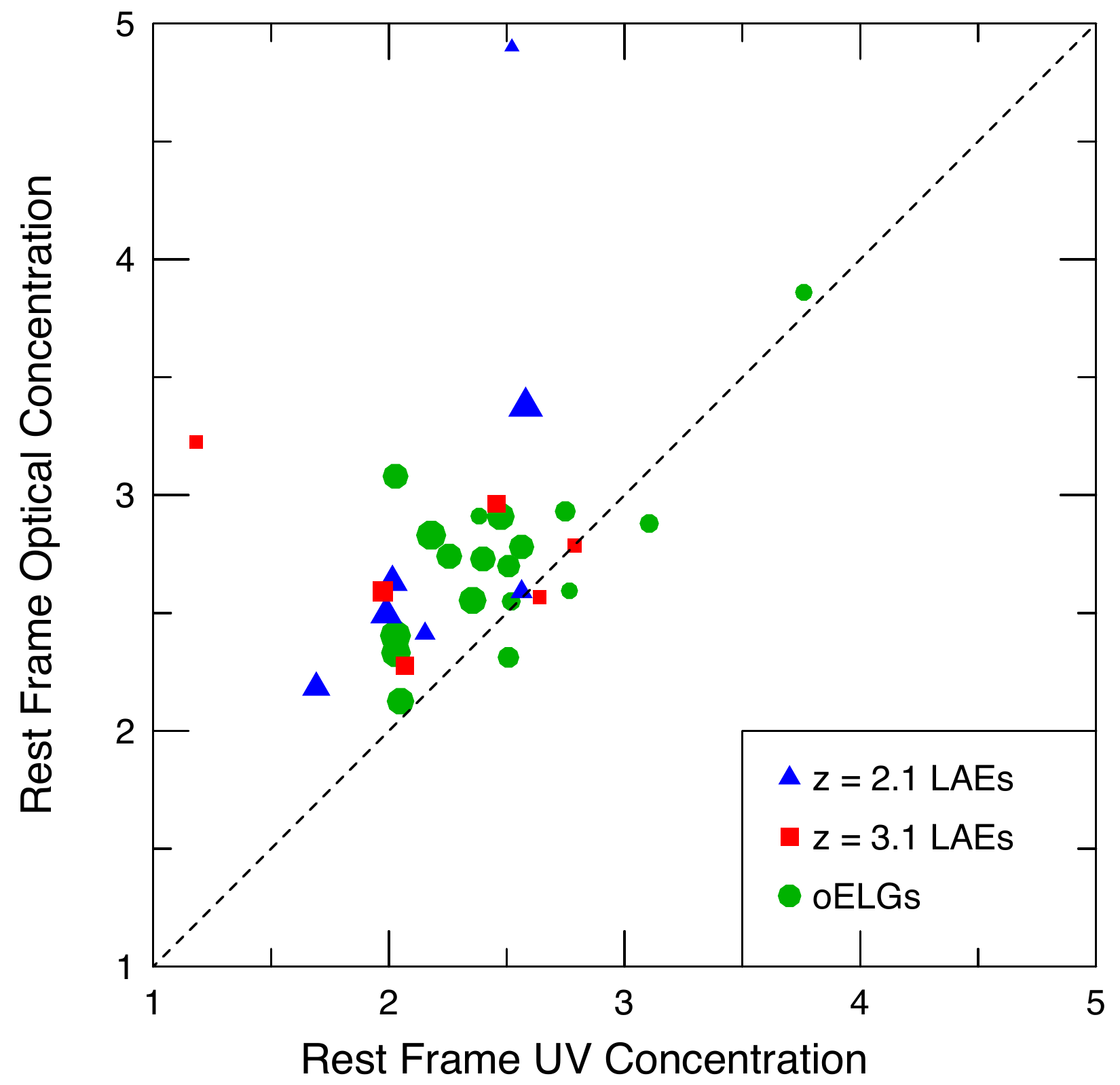}
\caption{The \citet{Kent85} concentration index measured in the rest-frame ultraviolet compare to 
that derived from the rest-frame optical for $2 \lesssim z \lesssim 3$ star-forming galaxies.  The
plotting scheme is the same as Figures \ref{fig:radius_mass}. The sizes of the points are proportional 
to how extended the galaxies appear on the sky.  Only those galaxies that have been resolved on 
both the F160W and F814W frames have been included in the figure.  All three samples of galaxies are 
more concentrated in the rest-frame optical than the rest-frame UV; this result is consistent with 
what is inferred from the galaxy's half-light radii. The discrepancy in the sizes and 
concentrations suggests an inside-out scenario for galaxy formation.
\label{fig:CvC}}
\end{figure}

This wavelength shift in the appearance of our $z \gtrsim 2$ emission-line galaxies is also
reflected in the systems' mean concentration index.   The majority of the galaxies in our 
dataset are unresolved on the F160W frames, and hence their concentrations are undefined.  
However, if we only consider those objects which are resolved, there is a 
systematic change in their structure with wavelength.  Specifically, as Figure~\ref{fig:CvC} illustrates,
both the oELGs and LAEs appear more concentrated in the rest-frame optical than in 
the UV by a median value of $\Delta C = 0.21$.  Moreover, this offset is not due to
issues associated with small, semi-resolved objects, as the effect is larger in the
more extended galaxies.   Thus, the behavior we observe in the local universe, with galaxies
appearing less concentrated in the ultraviolet, is mimicked at high-$z$.

Could those objects which appear larger in the rest-frame optical than in the UV be associated with
mergers?   To investigate this possibility, we used the CANDELS catalog to compute the surface density
of objects present on the images, and then increased this number by $\sim 15\%$ to 
account of possible sources that are present in our SExtractor catalog but below the CANDELS limit.
Based on this density, objects projected within $0 \farcs 6$ of any program galaxy have less than a 
$10\%$ chance of being a chance superposition.  Those objects with companions satisfying
this criterion are shown in Figure~\ref{fig:RvR} as open symbols.   As one can see from the
figure, close-by companions are principally associated with large systems:  although
$\sim 20\%$ of the $z \gtrsim 2$ star-forming galaxies studied in this program have neighbors
superposed within $0 \farcs 6$, six out of our eight largest systems ($r_e(\mathrm{UV}) \gtrsim 3$~kpc)
belong to this subset.    Less clear is the evidence for associating the  $r_e(\mathrm{Opt}) >
r_e(\mathrm{UV})$ galaxies with mergers:  just 1/3 of those systems have nearby 
companions.   Thus, it is possible that galaxy interactions may have an affect on the
rest-frame optical sizes of star-forming galaxies, but the signal is weak at best.

\subsection{Checking for Systematics}

Figures~\ref{fig:sizes}, \ref{fig:RvR}, and \ref{fig:kaplan_ecdf} all demonstrate that our $z \gtrsim 2$
star forming galaxies appear smaller at observer's frame infrared wavelengths than  
in the observer's frame optical.  This behavior is opposite to the effect produced by diffraction, and the
opposite of that expected from localized starbursts within the galaxies, where the UV may be emitted 
in just a few star forming regions.  But it is still possible that the offset is due to an instrumental 
systematic or a product of our reduction procedures.  To test for such effects, we undertook a series of 
experiments to explore the effects of image depths, binning, and wavelength on the results.

To test for a systematic error due to image depth, we added variable amounts of
Gaussian noise to our images, and re-measured the half-light radii of the galaxies as a 
function of this noise.  This analysis indicates that even when the signal-to-noise ratio is 
decreased by a factor of 30, there is no bias in our size measurements.  (The precision
of our estimated radii decreases, of course, as there is much greater scatter in the computed
half-light radii.  However, the values of $r_e$ remain unbiased.)   This result suggests that our 
conclusions are not being biased due to variations in the depths of the images.

The $0 \farcs 06$~per pixel image scale of the WFC3's IR camera is twice as large as that
for the ACS, which was used for all the other bandpasses in this study.  
To test whether this observational constraint has any affect on our results, we ran a $2 \times 2$
boxcar smoothing algorithm over our ACS F814W frames (to mimic the effect of the
drizzled images' correlated noise), binned the resultant image $2 \times 2$, and
remeasured the sizes of our galaxies.  The result is that the half-light estimates on the
more coarsely binned frames are $\sim 5\%$ {\it larger\/} than those on the original image.
(This result is true even if we do not smooth prior to binning.)    Thus, the effect of the
larger pixel size is opposite that needed to explain the offset between observing windows.
   
Finally, we can examine the behavior of $r_e$ with wavelength using solely the CCD data from
the ACS\null.   The ACS images range over 4000~\AA\ in the observed frame, 
and, for objects at $z \sim 2.1$, extend from the far UV ($\sim 1400$~\AA) to the very near 
UV ($\sim 3000$~\AA\null).  This bandpass may be just large enough to check for 
a wavelength dependence on galaxy size that is independent of the WFC3 instrument.

\begin{figure}[t]
\plotone{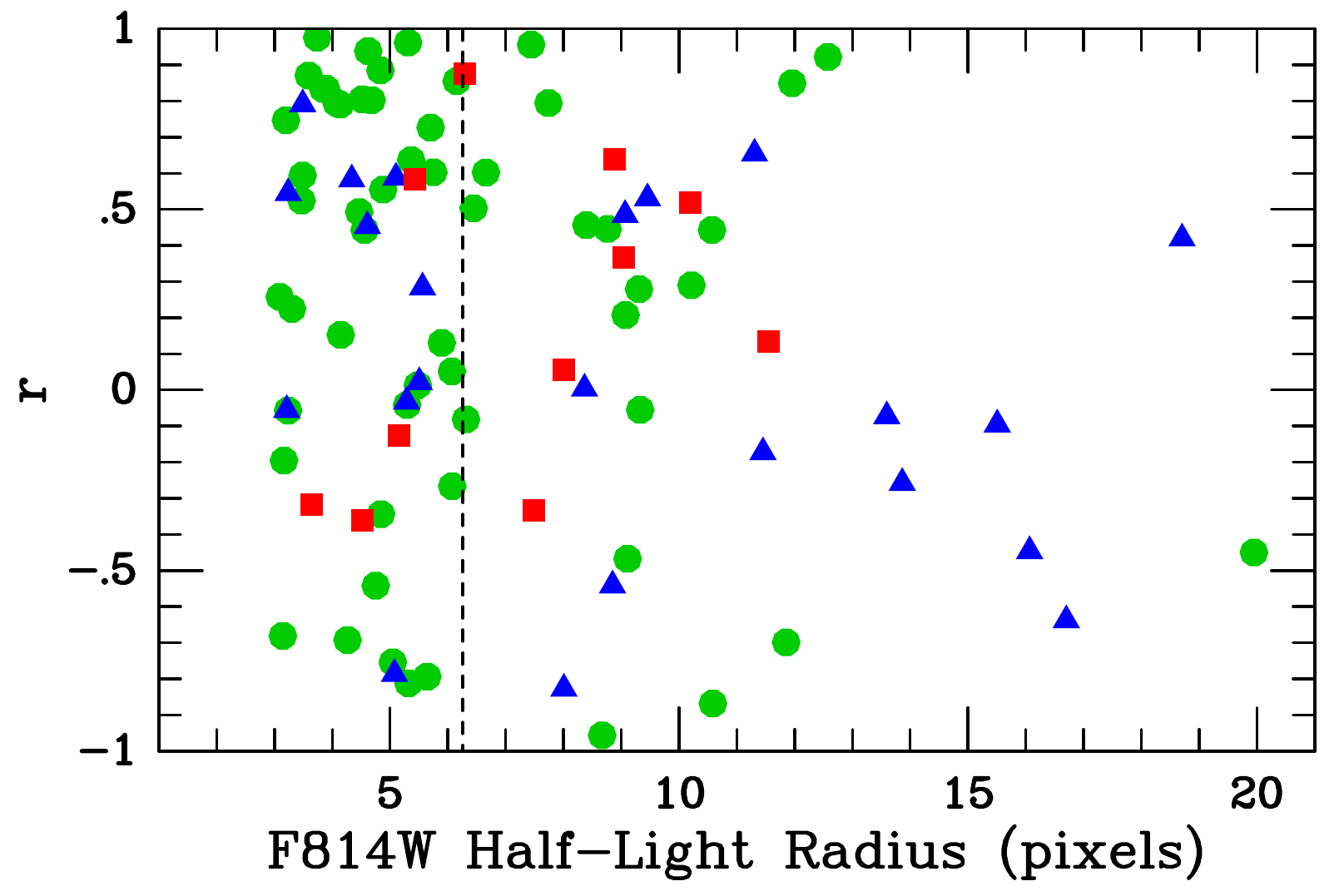}
\caption{The $y$-axis displays the Pearson correlation coefficient, $r$, derived from a regression 
of half-light radii measured in five of the ACS's filters (F435W, F606W, F775W, F814W, and 
F850LP)\null.  The $x$-axis shows the F814W half-light radius, plotted in pixels.
The dotted line is drawn at 1.5 times the value of $r_e$ one would derive for a 
point source. The plotting scheme is the same as Figure \ref{fig:radius_mass}.  (For $z=3.1$
sample, the F435W measurement was removed from the analysis.)  There is no strong 
evidence of any systematic behavior in the smallest objects, but in the larger systems, 
galaxy size is inversely correlated with the wavelength of observation. 
\label{fig:pearson}}
\end{figure}

Figure~\ref{fig:pearson} illustrates the results of this experiment by plotting the Pearson correlation
coefficient derived for a size versus wavelength regression against F814W half-light radius.  For the
oELGs and $z \sim 2.1$ LAEs, measurements through the ACS's
F435W, F606W, F775W, F814W, and F850LP filters all went into deriving the correlation coefficient,
while for the $z \sim 3.1$ LAEs, the bluest filter was omitted from the analysis (due to it lying blue ward of Lyman-$\alpha$).
For the smallest objects, 
i.e., those with radii only slightly larger than the frame's
point spread function, there is little evidence for systematic behavior:  if anything, the 
galaxies become larger with wavelength, as might be expected if the PSF were dominated 
by diffraction effects.   However, for the larger galaxies, there is clear evidence for an 
inverse correlation between the wavelength of observation and size.  Again, this result supports 
our conclusion that emission-line galaxies in the $2 \lesssim z \lesssim 3$ epoch are more 
extended in the rest-frame UV than they are in the optical. 

\section{DISCUSSION}
\label{sec:discussion}
There are three possible explanations for the shift in observed size for
$2 \lesssim z \lesssim 3$ star-forming galaxies.  Perhaps the simplest interpretation 
involves reddening due to dust.  In the local universe, the measured effective radius of a 
galaxy decreases significantly with increasing wavelength \citep[e.g.,][]{Kelvin+12, Vulcani+14, 
Lange+15}. This is mostly due to internal extinction: as one moves inward in a galaxy, the
surface density of dust increases, leading to an increase in obscuration, especially in the
ultraviolet.  The result is a shift in the balance of total flux towards larger radii, which artificially
increases the half-light radius \citep[e.g.,][]{Mollenhoff+06, Graham+08, Popescu+11}.
The same effect may be in play at the redshifts of our galaxies.  

To test for this possibility, we used the slope of the galaxies' UV spectral energy distribution ($\beta$),
which has been measured for all the oELGs in our sample \citep{Hagen+16}.  In general,
the stellar extinction in these galaxies is rather low:  if we adopt $\beta_0 = -2.25$ as
the expected UV slope from an unreddened starburst population \citep{Calzetti01}, then the median 
value of $\Delta \beta = 0.5$ translates into a stellar differential extinction of only
$E(B-V) = 0.11$ \citep{calzetti+00}.   Moreover, there is no significant correlation between
$\Delta \beta$ and $\Delta r_e$.  This strongly suggests that reddening is not responsible 
for the smaller values of optical $r_e$.

Alternatively, one might argue that our measurements are affected by the presence
of active galactic nuclei (AGN).  This explanation is extremely unlikely. \citet{Gronwall+07}
and \citet{Ciardullo+12} have tested for AGN activity in our samples of ECDF-S LAEs by
excluding objects detected in X-rays and by performing a deep X-ray stacking analysis on the 
objects.  Furthermore, \citet{Zeimann+14} reported that, while low-luminosity AGN could
be present in some of our oELGs, their contribution to the rest-frame UV continuum and the 
rest-frame optical emission lines must be very small.  Finally, any AGN that would be bright and 
unobscured enough to affect the distribution of optical light would most likely be even brighter in 
the UV, thereby shortening the half-light radii.  This effect would work against the trend we observe.

Finally, the shift of $r_e$ with wavelength may arise from a gradient in the stellar population. 
Inside-out models of galaxy formation predict that older stars should condense towards the 
centers of galaxies, perhaps in a bulge or pseudo-bulge \citep{vandenbosch+98, Dekel+09, 
Agertz+11}.  Inside-out formation has previously been observed in massive galaxies 
of the $z \sim 2$ epoch \citep[e.g.,][]{Patel+13, Lang+14}, and for $M \gtrsim 10^9 \, M_\odot$ 
galaxies at $z \sim 1$ \citep{Bond+14, vanderwel+14, Nelson+13, Nelson16}.  Here, we see
 that the process is also at work in $z \sim 2$ galaxies with much lower stellar 
masses, $7.5 \lesssim \log_{10} M_\star/M_{\odot} \lesssim 10$.  Our results suggest that even the
low-mass galaxies of the epoch are beginning to form a bulge or pseudo-bulge, with
older stars concentrated in the middle of the galaxy. This is unexpected, since
\cite{vanderwel+14} found that the effective size as a function of wavelength was itself a strong function
of mass, and the change was negligible for low-mass galaxies. However, the stellar mass limit of
\cite{vanderwel+14} is $\sim 10^9$, and so their analysis does not probe much of the mass range
analyzed here.

The next steps towards understanding the morphological properties of low-mass galaxies in
the $z \gtrsim 2$ universe are to vary the sample selection and broaden the redshift range. A proper
investigation of structure will then require the improved depth and spatial resolution of the 
Near-Infrared Camera (NIRcam) on the James Webb Space Telescope \citep[JWST;][]{JWST, nircam, 
nircam2}.  The near-infrared PSF of NIRcam will cover
$0 \farcs 05$, which corresponds to a $z \sim 2$ spatial scale of $0.5$ kpc.  
We look forward with anticipation to JWST's launch and science operations.

\acknowledgments

Support for program \#AR-13234 was provided by NASA through a 
grant from the Space Telescope Science Institute, which is operated by the Association of 
Universities for Research in Astronomy, Inc., under NASA contract NAS 5-26555.  
This work is based on observations made with the NASA/ESA Hubble Space Telescope
which is operated by the 
Association of Universities for Research in Astronomy, Inc., under NASA contract NAS5-26555.  
The data were obtained from the Hubble Legacy Archive, which is a collaboration between the 
Space Telescope Science Institute (STScI/NASA), the Space Telescope European 
Coordinating Facility (ST-ECF/ESA) and the Canadian Astronomy Data Centre 
(CADC/NRC/CSA\null).

The Institute for Gravitation and the Cosmos is supported by the Eberly College of Science 
and the Office of the Senior Vice President for Research at the Pennsylvania 
State University. This research has made use of NASA's Astrophysics Data System and the 
python packages \texttt{IPython}, \texttt{AstroPy}, 
\texttt{NumPy}, \texttt{SciPy}, \texttt{lifelines}, \texttt{pandas}, 
\texttt{matplotlib}, and \texttt{photutils}. \citep{ipython, astropy, numpy, scipy, lifelines, pandas, matplotlib, photutils}.



\clearpage


\renewcommand{\thefootnote}{\alph{footnote}}

\tablecaption{oELG Morphology Results \label{tbl:oelg}}
\begin{deluxetable}{crccccccc}
\tablehead{
&&&&&\multicolumn{2}{c}{Concentration} &\multicolumn{2}{c}{Half Light Radii (arcsec)} \\
\colhead{Field} &\colhead{ID} & \colhead{$\alpha(2000)$} &\colhead{$\delta(2000)$} 
&\colhead{$z$} &\colhead{F160W} &\colhead{F814W} &\colhead{F160W} 
&\colhead{F814W}  }
\startdata
30 &  272 & 53.07186 &$-27.82070$ &1.965 &2.73 &2.40  &$0.170 \pm 0.003$ &$0.272 \pm 0.001$ \\
32 & 6417 & 53.14214 &$-27.83269$ &2.153 &2.66 &1.67  &$<0.143$          &$0.193 \pm 0.001$ \\
34 & 2329 & 53.22949 &$-27.86480$ &2.171 &2.88 &3.10  &$0.348 \pm 0.010$ &$0.139 \pm 0.001$ \\
35 & 3033 & 53.13376 &$-27.80791$ &1.964 &3.61 &3.35  &$<0.143$          &$0.598 \pm 0.004$ \\
35 & 3485 & 53.14352 &$-27.79522$ &2.035 &3.86 &3.76  &$0.268 \pm 0.005$ &$0.099 \pm 0.001$ \\
36 &  140 & 53.17578 &$-27.81656$ &2.079 &2.87 &2.59  &$<0.143$          &$0.137 \pm 0.002$ \\
38 & 5800 & 53.13813 &$-27.86549$ &1.947 &3.60 &2.73  &$<0.143$          &$0.108 \pm 0.001$ \\
38 & 8270 & 53.13007 &$-27.84289$ &1.940 &3.02 &2.98  &$<0.143$          &$0.115 \pm 0.001$ \\
42 & 7854 & 53.18730 &$-27.89793$ &1.990 &2.82 &2.45  &$<0.143$          &$0.260 \pm 0.003$ \\
42 & 8611 & 53.18976 &$-27.89537$ &2.059 &2.65 &2.48  &$<0.143$          &$0.164 \pm 0.001$ \\
43 & 7274 & 53.09959 &$-27.93800$ &2.103 &2.49 &2.26  &$<0.143$          &$0.279 \pm 0.002$ \\
44 & 7544 & 53.16297 &$-27.91688$ &2.136 &2.94 &2.31  &$<0.143$          &$0.232 \pm 0.001$ \\
45 &  807 & 53.10760 &$-27.76926$ &2.076 &2.69 &2.79  &$<0.143$          &$0.171 \pm 0.002$ \\
45 &  950 & 53.11179 &$-27.76718$ &2.079 &2.75 &2.89  &$<0.143$          &$0.169 \pm 0.001$ \\
47 & 1069 & 53.05181 &$-27.84883$ &1.945 &2.55 &2.36  &$0.171 \pm 0.004$ &$0.318 \pm 0.002$ \\
47 & 3211 & 53.07551 &$-27.82831$ &2.034 &2.59 &2.73  &$<0.143$          &$0.182 \pm 0.001$ \\
48 &  366 & 53.10337 &$-27.84163$ &2.026 &3.24 &2.68  &$<0.143$          &$0.095 \pm 0.001$ \\
49 & 3237 & 53.16832 &$-27.87671$ &2.098 &3.08 &2.03  &$0.336 \pm 0.007$ &$0.273 \pm 0.001$ \\
49 & 3613 & 53.16614 &$-27.87466$ &2.131 &2.60 &2.55  &$<0.143$          &$0.122 \pm 0.001$ \\
49 & 7054 & 53.18917 &$-27.86309$ &1.970 &2.44 &2.25  &$<0.143$          &$0.159 \pm 0.001$ \\
50 & 6492 & 53.23087 &$-27.89474$ &1.939 &2.55 &2.52  &$0.391 \pm 0.029$ &$0.136 \pm 0.003$ \\
50 & 6991 & 53.24629 &$-27.88908$ &2.031 &2.70 &2.51  &$0.158 \pm 0.004$ &$0.223 \pm 0.001$ \\
50 & 7651 & 53.24427 &$-27.88427$ &2.318 &3.28 &2.07  &$<0.143$          &$0.252 \pm 0.005$ \\
51 & 2927 & 53.03610 &$-27.71394$ &2.075 &2.83 &2.18  &$0.226 \pm 0.004$ &$0.359 \pm 0.008$ \\
51 & 4972 & 53.01040 &$-27.71408$ &2.087 &2.33 &2.03  &$0.196 \pm 0.003$ &$0.356 \pm 0.004$ \\
52 & 3988 & 53.04419 &$-27.76833$ &2.198 &5.16 &2.62  &$<0.143$          &$0.151 \pm 0.003$ \\
53 & 3713 & 53.03542 &$-27.80838$ &2.012 &2.71 &2.92  &$<0.143$          &$0.096 \pm 0.001$ \\
53 & 7617 & 53.04000 &$-27.79442$ &2.040 &2.75 &2.98  &$<0.143$          &$0.145 \pm 0.001$ \\
53 &10334 & 53.02820 &$-27.77938$ &1.958 &3.96 &3.41  &$<0.143$          &$0.143 \pm 0.003$ \\
54 & 8192 & 53.08472 &$-27.86133$ &1.923 &2.78 &2.56  &$0.155 \pm 0.004$ &$0.263 \pm 0.001$ \\
55 & 3245 & 53.19357 &$-27.84351$ &2.009 &2.51 &2.30  &$<0.143$          &$0.185 \pm 0.001$ \\
55 & 6276 & 53.21752 &$-27.82591$ &2.211 &2.74 &3.11  &$<0.143$          &$0.112 \pm 0.001$ \\
55 & 6384 & 53.19595 &$-27.82481$ &2.090 &2.65 &2.67  &$<0.143$          &$0.140 \pm 0.001$ \\
55 & 6583 & 53.19229 &$-27.82294$ &2.032 &2.47 &2.45  &$<0.143$          &$0.145 \pm 0.001$ \\
55 & 7946 & 53.20340 &$-27.81601$ &1.999 &2.91 &2.47  &$0.177 \pm 0.003$ &$0.317 \pm 0.001$ \\
57 & 1058 & 53.18804 &$-27.74458$ &2.134 &2.80 &3.41  &$<0.143$          &$0.160 \pm 0.004$ \\
57 & 2412 & 53.19244 &$-27.73599$ &1.969 &2.40 &2.03  &$0.196 \pm 0.004$ &$0.377 \pm 0.003$ \\
57 & 2674 & 53.18128 &$-27.73417$ &1.946 &2.45 &2.35  &$<0.143$          &$0.146 \pm 0.001$ \\
57 & 3737 & 53.19071 &$-27.72948$ &2.250 &3.07 &4.68  &$<0.143$          &$0.134 \pm 0.003$ \\
57 & 4713 & 53.18109 &$-27.72681$ &2.078 &2.60 &2.58  &$<0.143$          &$0.182 \pm 0.002$ \\
57 & 4879 & 53.17942 &$-27.72628$ &2.221 &2.13 &2.05  &$0.186 \pm 0.004$ &$0.306 \pm 0.003$ \\
57 & 8227 & 53.18195 &$-27.71891$ &2.105 &2.90 &2.62  &$<0.143$          &$0.124 \pm 0.001$ \\
59 &  217 & 53.09352 &$-27.80971$ &2.324 &2.67 &2.57  &$<0.143$          &$0.124 \pm 0.001$ \\
59 &  951 & 53.09040 &$-27.80183$ &2.014 &2.91 &2.38  &$0.508 \pm 0.015$ &$0.104 \pm 0.001$ \\
59 & 1736 & 53.08912 &$-27.79275$ &2.255 &2.63 &2.69  &$<0.143$          &$0.105 \pm 0.001$ \\
59 & 2130 & 53.08884 &$-27.78168$ &1.947 &2.93 &2.75  &$0.218 \pm 0.004$ &$0.172 \pm 0.001$ \\
59 & 3424 & 53.11297 &$-27.77869$ &2.239 &2.74 &2.26  &$0.154 \pm 0.004$ &$0.280 \pm 0.008$ \\
60 &  247 & 53.14352 &$-27.79522$ &2.029 &3.86 &3.76  &$0.268 \pm 0.005$ &$0.099 \pm 0.001$ \\
60 & 1017 & 53.13915 &$-27.78615$ &1.987 &2.59 &2.77  &$0.338 \pm 0.014$ &$0.093 \pm 0.001$ \\
60 & 1362 & 53.12986 &$-27.78225$ &2.137 &2.50 &2.58  &$<0.143$          &$0.161 \pm 0.001$ \\
60 & 2437 & 53.14781 &$-27.77136$ &2.175 &2.92 &2.97  &$<0.143$          &$0.097 \pm 0.001$ \\
60 & 3248 & 53.14777 &$-27.76562$ &2.316 &2.31 &2.51  &$0.147 \pm 0.003$ &$0.189 \pm 0.001$ \\
63 & 1506 & 53.16881 &$-27.79694$ &1.992 &2.80 &3.59  &$<0.143$          &$0.128 \pm 0.001$ \\
63 & 1687 & 53.14832 &$-27.79594$ &2.030 &2.87 &2.70  &$<0.143$          &$0.117 \pm 0.002$ \\
63 & 1756 & 53.14352 &$-27.79522$ &2.031 &3.86 &3.76  &$0.268 \pm 0.005$ &$0.099 \pm 0.001$ \\
63 & 2814 & 53.13915 &$-27.78615$ &1.995 &2.59 &2.77  &$0.338 \pm 0.014$ &$0.093 \pm 0.001$ \\
63 & 3188 & 53.18222 &$-27.78331$ &2.081 &3.21 &\dots &$0.404 \pm 0.013$ &\dots             \\
63 & 3384 & 53.16746 &$-27.78183$ &2.069 &3.26 &2.75  &$<0.143$          &$0.159 \pm 0.003$ \\
63 & 3907 & 53.14889 &$-27.77750$ &2.073 &3.07 &2.58  &$<0.143$          &$0.094 \pm 0.001$ \\
63 & 5097 & 53.15381 &$-27.76730$ &2.317 &2.71 &2.84  &$<0.143$          &$0.177 \pm 0.001$ \\
63 & 5562 & 53.14781 &$-27.77136$ &2.177 &2.92 &2.97  &$<0.143$          &$0.097 \pm 0.001$ \\
\enddata
\end{deluxetable}

\begin{deluxetable}{rccccccc}
\tablecaption{LAE Morphology Results \label{tbl:lae}}
\tablehead{
&&&&\multicolumn{2}{c}{Concentration} &\multicolumn{2}{c}{Half Light Radii (arcsec)} \\
\colhead{ID} & \colhead{$\alpha(2000)$} &\colhead{$\delta(2000)$} &\colhead{$z$} 
&\colhead{F160W} &\colhead{F814W} &\colhead{F160W} &\colhead{F814W}  }
\startdata
 35 & 53.09592 &$-27.95504$ & 2.06 & 2.49 & 1.99 &$0.379 \pm 0.003$ &$0.501 \pm 0.010$ \\
 38 & 53.09076 &$-27.94879$ & 2.06 & 3.27 & 2.49 &$<0.143$          &$0.251 \pm 0.003$ \\
 40 & 53.12550 &$-27.94153$ & 2.06 & 2.18 & 1.69 &$0.220 \pm 0.002$ &$0.408 \pm 0.020$ \\
 50 & 53.12928 &$-27.91751$ & 2.06 & 3.71 & 2.47 &$<0.143$          &$0.167 \pm 0.003$ \\
 55 & 53.18977 &$-27.89537$ & 2.06 & 2.65 & 2.49 &$<0.143$          &$0.165 \pm 0.001$ \\
 60 & 53.06520 &$-27.87591$ & 2.06 & 2.64 & 2.54 &$<0.143$          &$0.096 \pm 0.001$ \\
 66 & 53.08463 &$-27.85105$ & 2.06 & 3.82 & 2.59 &$<0.143$          &$0.153 \pm 0.001$ \\
 68 & 53.04731 &$-27.84727$ & 2.06 & 2.63 & 2.02 &$0.276 \pm 0.006$ &$0.465 \pm 0.004$ \\
 69 & 53.12413 &$-27.84367$ & 2.06 & 2.59 & 2.56 &$0.171 \pm 0.005$ &$0.265 \pm 0.002$ \\
 70 & 53.04772 &$-27.84075$ & 2.06 & 4.90 & 2.52 &$0.461 \pm 0.007$ &$0.105 \pm 0.001$ \\
 75 & 53.16885 &$-27.82570$ & 2.06 & 2.87 & 1.79 &$<0.143$          &$0.284 \pm 0.018$ \\
 78 & 53.17580 &$-27.81655$ & 2.06 & 2.85 & 2.61 &$<0.143$          &$0.138 \pm 0.002$ \\
 83 & 53.17307 &$-27.80679$ & 2.06 & 3.88 & 2.74 &$<0.143$          &$0.097 \pm 0.002$ \\
 86 & 53.21541 &$-27.80223$ & 2.06 & 2.33 & 4.04 &$<0.143$          &$0.130 \pm 0.005$ \\
 90 & 53.13286 &$-27.79795$ & 2.06 & 5.21 & 1.81 &$<0.143$          &$0.339 \pm 0.010$ \\
 98 & 53.16750 &$-27.78188$ & 2.06 & 2.64 & 2.58 &$<0.143$          &$0.159 \pm 0.003$ \\
115 & 53.01718 &$-27.75078$ & 2.06 & 3.17 & 3.09 &$<0.143$          &$0.152 \pm 0.002$ \\
126 & 53.01668 &$-27.73100$ & 2.06 & 3.38 & 2.58 &$0.265 \pm 0.014$ &$0.561 \pm 0.011$ \\
130 & 53.18053 &$-27.72655$ & 2.06 & 2.41 & 2.15 &$0.156 \pm 0.006$ &$0.240 \pm 0.006$ \\
141 & 53.05732 &$-27.71680$ & 2.06 & 6.24 & 1.93 &$<0.143$          &$0.272 \pm 0.004$ \\
144 & 53.03030 &$-27.69660$ & 2.06 &\dots & 1.78 &\dots             &$0.482 \pm 0.019$ \\
156 & 53.14160 &$-27.67207$ & 2.06 & 2.82 & 1.93 &$<0.143$          &$0.416 \pm 0.006$ \\
158 & 53.13110 &$-27.66999$ & 2.06 &\dots & 2.79 &\dots             &$0.344 \pm 0.008$ \\
  4 & 53.07833 &$-27.71338$ & 3.12 & 2.59 & 1.97 &$0.173 \pm 0.002$ &$0.347 \pm 0.013$ \\
  6 & 53.21955 &$-27.80259$ & 3.12 & 2.79 & 2.79 &$0.531 \pm 0.042$ &$0.163 \pm 0.002$ \\
 11 & 53.11215 &$-27.69116$ & 3.12 & 3.22 & 1.18 &$0.581 \pm 0.006$ &$0.136 \pm 0.001$ \\
 16 & 53.05533 &$-27.72499$ & 3.12 &\dots & 3.54 &\dots             &$0.252 \pm 0.005$ \\
 22 & 53.16187 &$-27.69557$ & 3.12 & 2.55 & 2.39 &$<0.143$          &$0.225 \pm 0.003$ \\
 25 & 53.16994 &$-27.76833$ & 3.12 & 2.96 & 2.46 &$0.157 \pm 0.005$ &$0.271 \pm 0.001$ \\
 44 & 53.06578 &$-27.73618$ & 3.12 & 2.28 & 2.07 &$0.144 \pm 0.005$ &$0.267 \pm 0.005$ \\
 56 & 53.14304 &$-27.79988$ & 3.12 & 2.57 & 2.64 &$0.318 \pm 0.014$ &$0.155 \pm 0.004$ \\
 59 & 53.13857 &$-27.85766$ & 3.12 & 2.80 & 2.50 &$<0.143$          &$0.240 \pm 0.004$ \\
 93 & 53.16438 &$-27.74711$ & 3.12 & 2.69 & 2.84 &$<0.143$          &$0.189 \pm 0.001$ \\
 94 & 53.03884 &$-27.73179$ & 3.12 &\dots & 2.35 &\dots             &$0.109 \pm 0.001$ \\
125 & 53.16256 &$-27.77286$ & 3.12 & 2.01 & 2.18 &$<0.143$          &$0.306 \pm 0.006$ \\
\enddata
\end{deluxetable}


\end{document}